\documentstyle[12pt,epsfig]{article}
\def\beq{\begin{equation}}
\def\eeq{\end{equation}}
\def\bea{\begin{eqnarray}}
\def\eea{\end{eqnarray}}
\def\non{\nonumber}

\def\frh{\frac{1}{2}}

\begin{document}

\begin{center}
{\Large \bf \sf Pseudo-hermitian interaction between an oscillator
and a spin-$\frh$ particle in the external magnetic field  }

\vspace{1.3cm}

{\sf  Bhabani Prasad Mandal \footnote{e-mail address:
bpm@bose.res.in}}

\bigskip

{\em Department of Physics, Banaras Hindu University, \\
Varanasi-221005, India}

\bigskip
\bigskip

\noindent {\bf Abstract}

\end{center}

 We consider a spin-$\frh$ particle in the external magnetic field
 which couples to a harmonic oscillator through some pseudo-hermitian
 interaction. We find that the energy eigenvalues for this system
 are real even though
the interaction is not PT invariant.


\noindent \section{Introduction}
\renewcommand{\theequation}{1.{\arabic{equation}}}
\setcounter{equation}{0}

\medskip

In the last few years the study of some nonhermitian Hamiltonian
with real spectrum have given rise to a growing interest in the
literature. This was mainly initiated by Bender and Boettcher's
observation that with properly defined boundary conditions the
spectrum of the Hamiltonian $ H=p^2 +x^2 (ix)^\nu, \ \ \ (\nu\geq
0)$ is real, positive and discrete. The reality of the spectrum is
a consequence of unbroken PT [ combined parity (P) and time
reversal (T) ] invariance of the Hamiltonian i.e. $[H, PT]=0$ \
\cite{ben,pt1}. However pairs of complex conjugate eigenvalues
appear when the PT symmetry is broken spontaneously. This is also
illustrated nicely with the help of a nonhermitian but PT
invariant potential with quasi-exactly solvable eigenvalues
\cite{pt3} .

This surprising result attracts lot of interest in last few years
and many other such nonhermitian but PT symmetric systems, mostly
for one particle in one space dimension have been investigated
\cite{ben1}- \cite{kha}. Validity of these results have also been
tested for the cases of nonhermitian extension of some exactly
solvable many particle quantum systems in one dimension
\cite{bm} -\cite{bn}. Nonhermitian extension of some field
theoretic models has been considered in Refs. \cite{ben2,ka}.

However to develop a consistent quantum theory for these
nonhermitian Hamiltonians  one encounters  serve difficulties
\cite{jap,ddt}. Firstly, the eigenstates of PT symmetric
nonhermitian Hamiltonians with real eigenvalues only do not
satisfy standard completeness relations. More importantly  if one
takes the natural inner product associated with PT-symmetric
system  as $$ (f ,g) = \int d^4x [PT f(x)] g(x), $$ then the half
of the energy eigenstates have negative norms which makes  it
difficult to maintain the familiar probabilistic interpretation of
quantum theory. Recently Bender and coworkers have found a new
symmetry , C, inherent in all such Hamiltonian with unbroken PT
symmetry\cite{pt1,ben1}. This allows to introduce an inner product
structure associated with CPT conjugation for which the norms of
the quantum states are positive definite and one gets usual
completeness relation. As a result the Hamiltonian and its
eigenstates can be extended  to complex domain so that the
associated eigenvalues are real and underlying dynamics is
unitary.

In an another approach Mostafazadeh \cite{ali,ali1} has shown that
the reality of spectrum of nonhermitian Hamiltonian is due to so
called pseudo-hermiticity properties of the Hamiltonian. A
Hamiltonian is called $\eta $ pseudo-hermitian if it satisfies the
relation \beq \eta H \eta^{-1} =H^\dag , \eeq where $\eta$ is a
linear hermitian operator. All PT symmetric nonhermitian
Hamiltonian are pseudo-hermitian and these consist a subclass of
pseudo-hermitian Hamiltonian. All the observations of PT symmetric
nonhermitian Hamiltonian can be explained nicely in this approach.

The purpose of this letter is to consider an example of
nonhermitian Hamiltonian which is not PT invariant but
pseudo-hermitian and study the different properties of such
system. With this aim we consider a system consisting of a spin
half particle in the external magnetic field coupled to an
oscillator via nonhermitian interaction. We find that the spectrum
is real even though the interaction term is not PT symmetric. The
explicit PT asymmetric system  has also been considered recently
\cite{23}.

Here is the plan of the paper. In section II we will discuss the
Hamiltonian of the system and its symmetries. We will find the
energy eigenvalues  and corresponding  eigenfuctions explicitly for this system in
section III. Section IV is kept for concluding remarks.

 \vspace{1cm}

\noindent \section{ The Model}
\renewcommand{\theequation}{2.{\arabic{equation}}}
\setcounter{equation}{0}

We consider a system of a spin $\frh$ particle in the external
magnetic field, $\vec{B}$ coupled to an oscillator through some
nonhermitian interaction  described by the Hamiltonian \beq H=
\mu\vec{\sigma}\cdot\vec{B}+\hbar\omega a^\dag a + \rho(\sigma_+a
-\sigma_- a^\dag) \label{h} .\eeq Here $\vec{\sigma}$'s are Pauli
matrices, $\rho$ is some arbitrary real parameter and
$\sigma_{\pm} \equiv\frac{1}{2}[ \sigma_x \pm i \sigma_y]$ are
spin projection operators. $ a, a^\dag$ are usual creation and
annihilation operator for the oscillator states and defined as
\beq a = \frac{p-im\omega x}{\sqrt{2m\omega\hbar}},\ \ \
a^\dag=\frac{p+im\omega x}{\sqrt{2m\omega\hbar}}\label{aa}, \eeq
with \beq a|n> = \sqrt{n}|n-1> ,\ \ \mbox{and}\ \ a^\dag|n> =
\sqrt{n+1}|n+1> \label{aa1},\eeq where the notation $|n> $ for
number eigenvectors for the oscillator has been adopted.

For the sake of simplicity we can choose the external magnetic
field in z-direction, $ \vec{B}= B_0 \hat{z}$ and the Hamiltonian
for the system as given in Eq.(\ref{h}) is reduced to , \beq
H=\frac{\epsilon}{2}\sigma_z+\hbar\omega a^\dag a + \rho(\sigma_+a
-\sigma_- a^\dag) \label{h1}, \eeq where $\epsilon=2\mu B_0$. This
system can also be thought of a two level system   coupled to an
oscillator where $\epsilon$ is the splitting between the levels.
Note that this Hamiltinian is not hermitian as, \bea
H^\dag&=&\frac{\epsilon}{2}\sigma_{z}+\hbar\omega a^\dag a -
\rho(\sigma_+a -\sigma_- a^\dag), \nonumber
\\ &\neq & H, \label{hh}\eea as $ \sigma_{\pm}^{\dag}=\sigma_{\mp}.$
  Under parity transformation [i.e. $ x\longrightarrow -x;\  p\longrightarrow-p$]
   both $\vec{\sigma}$ and $ \vec{B}$
do not change sign as both are axial vectors but as it clear from
the Eq (\ref{aa}) that both the creation and annihilation
operators change sign. \bea && P\vec{\sigma}P^{-1} =
\vec{\sigma},\nonumber \\&& P\vec{B}P^{-1}= \vec{B},\non \\
&&PaP^{-1} = -a ,\non \\ && Pa^\dag P^{-1} = -a^\dag
\label{p}.\eea
 Note the interaction term
of the Hamiltonian in Eq. (\ref{h})  changes sign under parity
operation. The time reversal  operator for the system of spin half
particles is $T= -i\sigma_y K$ where K is complex conjugation
operator. We note the changes of following quantities under time
reversal transformation as, \bea && T\vec{\sigma}T^{-1} =
-\vec{\sigma}\nonumber, \\ && T\vec{B}T^{-1} = \vec{B}, \nonumber
\\ && T\sigma_{\pm} T^{-1} = -\sigma_{\mp}, \nonumber \\ && TaT^{-1} =
-a, \nonumber \\ && Ta^\dag T^{-1} = -a^\dag \label{t}.\eea We
have considered the magnetic field as the external element in our
system and it does not change sign under time reversal operation.
However one can consider magnetic field in other way also, when it
changes sign under time reversal as the current producing magnetic
field is  reversed under time reversal. The results of this paper
are  same  in both cases. From Eqs. (\ref{p}) and (\ref{t}) we can
see that the Hamiltonia in Eq. (\ref{h}) is not PT symmetric, \bea
 PT\  H\ {(PT)}^{-1}= &-&\frac{\epsilon}{2}\sigma_{z}+\hbar\omega
a^\dag a + \rho(\sigma_+a^\dag -\sigma_- a),\nonumber \\ &\neq & H
\label{hpt}.\eea

However this Hamiltonian is $\sigma_z$-pseudo-hermitian \bea
\sigma_{z}H \sigma_{z}^{-1}
&=&\frac{\epsilon}{2}\sigma_z+\hbar\omega a^\dag a +
\rho(\sigma_z\sigma_+\sigma_za -\sigma_z\sigma_-\sigma_za^\dag)
,\non \\&=&\frac{\epsilon}{2}\sigma_z+\hbar\omega a^\dag a
-\rho(\sigma_{+}a-\sigma_-a^\dag),\non \\ &=& H^\dag
\label{psh}.\eea Here we have used the relations
$\sigma_z\sigma_{\pm}\sigma_z = -\sigma_{\pm}$. In case of $\eta-$
pseudo-hermitian Hamiltonian the choice of the operator $\eta$ is
not unique \cite{ali}. Therefore we look for whether our
Hamiltonian is pseudo-hermitian with respect to any other
operator. Indeed it is also pseudo-hermitian with respect to
parity operator as, \bea PHP^{-1} &=&
\frac{\epsilon}{2}P\sigma_zP^{-1} +\hbar\omega P
a^{\dag}aP^{-1}+\rho(P\sigma_+ aP^{-1} -P\sigma_-a^{\dag}P^{-1})
,\non
\\ &=&\frac{\epsilon}{2}\sigma_z +\hbar\omega
a^{\dag}a-\rho(\sigma_+ a -\sigma_-a^{\dag} ),\non\\ &=&
H^\dag.\label{psh1}\eea Finally we found a symmetry of our
Hamiltonian . It is invariant under the symmetry generated by the
combined operator, $P\sigma_z$ i.e. \beq [H,P\sigma_z]=0.
\label{pz}\eeq However it not surprising as
 it is
shown in Ref.\cite{ali} that if a Hamiltonian is
pseudo-hermitian with respect to two different operator $\eta_1
,\eta_2 $ then the system is symmetric under the transformation
generated by $\eta_2^{-1}\eta_1$.

\noindent\section{ The solutions}
\renewcommand{\theequation}{3.{\arabic{equation}}}
\setcounter{equation}{0}

To find the energy eigenvalues and corresponding  eigenvectors of
the system described by the  Hamiltonian in the Eq. (\ref{h1}) we
adopt the  notation for the state as, $|n,\frac{1}{2}m_s>$ where
$n$ is eigenvalue for the number operator $a^{\dag}a$ i.e.
$a^{\dag}a|n> = n|n>$ and $m_s =\pm 1$ are the eigenvalues of the
operator $\sigma_{z}$ i.e. $
\sigma_{z}|\frac{1}{2}m_s>=m_s|\frac{1}{2}m_s>$. It is readily
seen that $|0,-\frac{1}{2}>$ is a ground state  of the Hamiltonian
with eigenvalue $-\frac{\epsilon}{2}$ and it is non-degenerate.
\beq H|0,-\frac{1}{2}>
=-\frac{\epsilon}{2}|0,-\frac{1}{2}>\label{gs}.\eeq  Note that the
projection operators $\sigma_{\pm}$ have the following usual
properties when they act on the state $|n,\pm\frac{1}{2}>$, \bea
\sigma_{+}|n,\frac{1}{2}>=0; &\ \ \ \ \
&\sigma_{+}|n,-\frac{1}{2}>=|n,\frac{1}{2}>, \non
\\ \sigma_{-}|n,-\frac{1}{2}>=0; &\ \ \ \ \
&\sigma_{-}|n,\frac{1}{2}>=|n,-\frac{1}{2}>\label{sig}.\eea  We
observe that the next possible states $|0,\frac{1}{2}> $ is not a
eigenstate of the Hamiltonian, \beq H|0,\frac{1}{2}>
=\frac{\epsilon}{2}|0,\frac{1}{2}>-\rho|1-\frac{1}{2}>
\label{1}.\eeq However this state along with the state
$|1,-\frac{1}{2}> $ close under the action of the Hamiltonian and
form a invariant subspace in the space of states  as, \beq H|1,
-\frac{1}{2}> = (\hbar\omega-\frac{\epsilon}{2})|1,-\frac{1}{2}>
+\rho|0,\frac{1}{2}>\label{2} .\eeq First two excited states
belong to this sector spanned by these two states,
$|0,\frac{1}{2}> $ and $|1,-\frac{1}{2}>$ wherein the Hamiltonian
matrix is given by\footnote{ Similar two by two matrix Hamiltonian
is also considered in ref. \cite{24} for a completely different
system.}
\[H_{1}=\left[ \begin{array}{cc}\frac{\epsilon}{2}& \rho
\\-\rho&-\frac{\epsilon}{2}+\hbar\omega\end{array}\right ].\]
The eigenvalues of this Hamiltonian matrix are given by
$\lambda_{1}^{I,II}=\frac{1}{2}\left [
\hbar\omega\pm\sqrt{(\hbar\omega-\epsilon)^2-4\rho^2}\right ].$
Note these eigenvalues are real provided
$|\hbar\omega-\epsilon|\geq 2\rho$. Putting
$2\rho=(\hbar\omega-\epsilon)\sin\theta_{1} $ we find the
eigenvectors corresponding to this doublet are \bea |\Psi_1^I> &=&
\cos\frac{\theta_{1}}{2}|0,\frac{1}{2}>
+\sin\frac{\theta_{1}}{2}|1,-\frac{1}{2}>,\ \ \  \mbox{for}\
\lambda_1^{I}=
\frac{\hbar\omega}{2}(1+\cos\theta_{1})-\frac{\epsilon}{2}\cos\theta_{1},\non\\
|\Psi_1^{II}> &=& \sin\frac{\theta_{1}}{2}|0,\frac{1}{2}>
+\cos\frac{\theta_{1}}{2}|1,-\frac{1}{2}>,\ \ \  \mbox{for}\
\lambda_2^{II}=
\frac{\hbar\omega}{2}(1-\cos\theta_{1})+\frac{\epsilon}{2}\cos\theta_{1}.
\nonumber \\ \label{11}\eea It may be observed that these two
states are not orthogonal to each other nor do they have to be as
$H\neq H^{\dag}.$ To find the next excited states we have to
consider next invariant subspace. It can be easily checked that
next invariant subspace is spanned by the
vectors,$|1,\frac{1}{2}>, |2,-\frac{1}{2}>$ and the eigenvalues
and eigenvectors for this doublet can be obtained following the
same method.

The result is easily generalized to the sector spanned by
$|n,\frac{1}{2}> $ and $ |n+1,-\frac{1}{2}>$ wherein the
Hamiltonian matrix is given by,\[H_{n+1}=\left[
\begin{array}{cc}\frac{\epsilon}{2}+n\hbar\omega & \rho\sqrt{n+1}
\\-\rho\sqrt{n+1}&-\frac{\epsilon}{2}+(n+1)\hbar\omega \end{array}\right ].\]
Now we have the eigenvalues of this Hamiltonian matrix are given
by \beq\lambda_{n+1}^{I,II}=\frac{1}{2}\left[(2n+1)\hbar\omega\pm
\sqrt{(\hbar\omega-\epsilon)^2-4\rho^2(n+1)}
\right]\label{ev}.\eeq These eigenvalues are real provided
$|\hbar\omega-\epsilon|\geq 2\rho\sqrt{n+1}$. Now putting
$2\rho\sqrt{n+1}=(\hbar\omega-\epsilon)\sin\theta_{n+1}, $ we find
the eigenvectors corresponding to this doublet are \bea
|\Psi_{n+1}^I> &=& \cos\frac{\theta_{n+1}}{2}|n,\frac{1}{2}>
+\sin\frac{\theta_{n+1}}{2}|n+1,-\frac{1}{2}>,\non\\ &&
\mbox{for}\ \lambda_{n+1}^I=
\frac{\hbar\omega}{2}(2n+1+\cos\theta_{n+1})-\frac{\epsilon}{2}\cos\theta_{n+1},\non
\\ |\Psi_{n+1}^{II}> &=& \sin\frac{\theta_{n+1}}{2}|n,\frac{1}{2}>
+\cos\frac{\theta_{n+1}}{2}|n+1,-\frac{1}{2}>,\non \\
&&\mbox{for}\ \lambda_{n+1}^{II}=
\frac{\hbar\omega}{2}(2n+1-\cos\theta_{n+1})+\frac{\epsilon}{2}\cos\theta_{n+1}
\label{n+1}.\eea We observe that these eigenstates are also
eigenstate of the operator, $P\sigma_z$ \beq P\sigma_z
|\Psi_{n+1}^{I,II}> = (-1)^n|\Psi_{n+1}^{I,II}> \label{pz1}, \eeq
as $P|n, \pm\frac{1}{2}> = (-1)^n |n,\pm\frac{1}{2}>$ and
$\sigma_{z}|n,\pm\frac{1}{2}> =\pm|n,\pm\frac{1}{2}>$. Thus we
have real eigenvalues when the symmetry is not broken. In the
regime $ |\hbar\omega-\epsilon|<2\rho\sqrt{n+1}$ the eigenvalues
for a particular doublet become complex conjugate. For $n+1$ th
doublet, the complex eigenvalues are $\frac{1}{2}\left
[\hbar\omega(2n+1)+\frac{i}{2}\sqrt{4\rho^{2}(n+1)-(\hbar\omega-\epsilon)^2}\right
].$

The Hamiltonian of the system has a complete set of biorthonormal
eigenvectors. To see this we find the eigenvalues and eigenvectors
of $H^{\dag}$ as,\bea |\Phi_{n+1}^I> &=&
\cos\frac{\theta_{n+1}}{2}|n,\frac{1}{2}>
-\sin\frac{\theta_{n+1}}{2}|n+1,-\frac{1}{2}>,\non\\ &&
\mbox{for}\ \tilde{\lambda}_{n+1}^I=
\frac{\hbar\omega}{2}(2n+1+\cos\theta_{n+1})-\frac{\epsilon}{2}\cos\theta_{n+1},\non
\\ |\Phi_{n+1}^{II}> &=&
-\sin\frac{\theta_{n+1}}{2}|n,\frac{1}{2}>
+\cos\frac{\theta_{n+1}}{2}|n+1,-\frac{1}{2}>,\non \\
&&\mbox{for}\ \tilde{\lambda}_{n+1}^2=
\frac{\hbar\omega}{2}(2n+1-\cos\theta_{n+1})+\frac{\epsilon}{2}\cos\theta_{n+1}
\label{n+2}.\eea  From the Eqs. (\ref{n+1} ) and (\ref{n+2}) one
can easily check that \bea &&<\Psi_n^{i}|\Phi_{m}^{j}> =
\delta_{nm}\delta_{ij}\ \ \ \mbox{ where } i,j = I \mbox{  or  }
II,
\\ &&\sum_{n}\sum_{i}|\Psi_{n}^{i}><\Phi_{n}^{i}| =
\sum_{n}\sum_{i}|\Phi_{n}^{i}><\Psi_{n}^{i}| =\mathbf{1}
\label{bi}, \eea modulo a constant scaling of the states.

Now we observe that  one state in each doublet has $P\sigma_{z}$
pseudo norm negative  \beq <\Psi_n^{i}|P\sigma_{z}|\Psi_m^i>
=\pm(-1)^n \delta_{mn}\ \  \mbox{for}\  i= I, II .\eeq  This fact
is similar to the fact in all  PT-symmetric nonhermitian systems
where the PT-norms for half of the states are negative. The system
we have considered is invariant under the the symmetry generated
by the combined operator $P\sigma_Z$ and we have $P\sigma_z$ norms
for  half of the states are negative. Now following the work in
ref [2] we introduce the extra symmetry, C, connected with the
equal numbers of positive and negative norms. The operator C is
the observable that represents the measurement of signature of
$P\sigma_{z}$ norm of a state. Then $PC\sigma_{z}$ norms for all
the states are positive and definite. Thus if we introduce the
inner product associated with our system as,

\begin{equation} (f,g) = \int d^4 x[P\sigma_z C f(x) ]g(x)
\label{nin},\end{equation}
 then all the states become orthonormal.

\noindent\section{Conclusion}
\renewcommand{\theequation}{4.{\arabic{equation}}}
\setcounter{equation}{0}

We consider a system of a spin half particle in external magnetic field
coupled to an oscillator through nonhermitian interaction. The
Hamiltonian for this system is not PT-symmetric but it is
 pseudo-hermitian with respect to two different operators, $P$
and $\sigma_{z}$ and hence symmetric under the transformations
generated by the operator $P\sigma_z$. We found that, except the ground state,
all other states occurs in doublet due to the interaction of
oscillator with two levels system. The energy eigenvalues corresponding to
 all these
states are real when the symmetry $P\sigma_{z}$ is unbroken. We
have also shown that the Hamiltonian of the system has a complete
set of biorthonormal eigenvectors. However with the conventional
definition of scalar product the eigenstates corresponding to a
particular doublet are not orthogonal to each other, but they are
orthogonal to all other eigenstates corresponding to other
doublets. The eigenstates of a particular doublet are orthogonal
to each other only when, $\theta_{n}=n\pi ,$ i.e. $\rho=0$. That
is the situation when the nonhermitian interaction drops. However
when the $P\sigma_z$ symmetry is broken spontaneously the energy
eigenvalues are complex and  all the eigenstates are orthogonal to
each other. This implies that when energy eigenvalues are
observable(real), it is not possible to have all the states
orthogonal to each other with the conventional definition of
scalar product, on the other hand all the eigenstates satisfy
 orthonormality condition when eigenvalues (complex) are  not observable.
This result is not surprising as the system is nonhermitian
\cite{jap}. However all the states satisfy orthonormality with
respect to the new definition of inner product in Eq. \ref{nin}.


\begin{thebibliography}{99}

\bibitem{ben}C.M. Bender and S. Boettcher, {\em Phys. Rev. Lett.}
{\bf 80} (1998) 5243; {\em J. Phys. } {\bf A 31} (1998) L273; C.M.
Bender, S. Boettcher and P.N. Meisinger, {\em J. Math. Phys.} {\bf
40} (1999) 2210.
\bibitem{pt1}C.M. Bender and S. Boettcher, {\em Phys. Rev. Lett.}
 {\bf 89} (2002) 270401-1;

\bibitem{pt3} A. Khare and B. P. Mandal,
{\em Phys. Lett. } {\bf A 272} (2000) 53.

\bibitem{ben1} C.M. Bender,J. Brod, A. Refig and M. Reuter,
quant-ph/0402026.

\bibitem{ben2} C.M. Bender, D.C. Brody and H. F. Jones, hep-th/0402183;
hep-th/0402011.

\bibitem{ka} K. A. Milton, {\em Czech J. Phys.} {\bf 54} (2004)
1069.


\bibitem{ali} A. Mostafazadeh, {\em J. Math Phys.}
 {\bf 43 } (2002) 205; {\bf 43} (2002) 2814; {\bf 43} (2002) 3944.

\bibitem{ali1} A. Mostafazadeh, {\em Nucl. Phys. B} {\bf 640} (2002) 419;
{\em J. Math. Phys.} {\bf 44} (2003) 974; quant-ph/0307059; quant-ph/0404025;
quant-ph/0304080.

\bibitem{wei} S. Weigert, quant-ph/0209054 and quant-ph/0306040.

\bibitem{jap} G. Japaridze, {\em J. Phys. A} {\bf 35} (2002),
1709.

\bibitem{ahm} Z. Ahmed, {\em Phys. Lett. A} {\bf 294} (2002), 287.

\bibitem{zno} M. Znojil, math-ph/0104012, quant-ph/0303122 \&
math-ph/0403033.

\bibitem{ddt} P. Dorey, C. Dunning and R. Tateo,
{\em J. Phys.} {\bf A 34} (2001) 5679.

\bibitem{bag} B.Bagchi, C. Quesne, {\em Phys. Lett.} {\bf A273}
(2000) 256.

\bibitem{pan} S. Rajvjani, A.K. kapoor and P. K. Panigrahi,
quant-ph/0403054.

\bibitem{anj} A. Sinha, G. Levai and P. Roy
quant-ph/0401064.

\bibitem{kha} A. Khare and U. Sukhatme, quant-ph/0402106

\bibitem{bm}B. Basu-Mallick and B.P. Mandal,
{\em Phys. Lett. } {\bf A 284} (2001) 231.

\bibitem{biru} B. Basu-Mallick, {\em Int. J. of Mod. Phys. B},  {\bf 16} (2002)
1875.

\bibitem{sca} B. Basu-Mallick, T. Bhattacharyya  A. Kundu, and B. P. Mandal
{\em Czech. J. Phys } {\bf 54} (2004) 5.

\bibitem{sca1} B. Basu-Mallick, T. Bhattacharyya  and B. P. Mandal,
nlin.SI/0405068 [To appear in IJMPA, 2004].



\bibitem{bn} Y. Brihaye and A. Ninimahazwe hep-th/0311081

\bibitem{23} E. Caliceti, F. Cannata, M. Znojil \& A. Ventura math-ph/0406031 .

\bibitem{24} A. Mostafazadeh, quant-ph/0310164.
\end{thebibliography}
\end{document}